\PassOptionsToPackage{cmyk,table}{xcolor}
\documentclass[conference,pbalance]{iaria}
\pdfoutput=1

\usepackage[babel=true,english=american]{csquotes}
\usepackage[english]{babel}
\usepackage[shortcuts]{extdash}
\usepackage[alwaysadjust]{paralist}
\usepackage[caption=false,font=footnotesize]{subfig}
\usepackage[
  babel=true,
  expansion=alltext,
  protrusion=alltext-nott,
  nopatch=eqnum,
  final
]{microtype}

\usepackage{booktabs}
\usepackage{multirow}
\usepackage{amsmath,amssymb}
\usepackage{graphicx}
\usepackage{tikz}
\usetikzlibrary{shapes.geometric,arrows.meta,positioning,fit}
\usepackage{xcolor}
\usepackage{enumitem}
\usepackage{xurl}

\graphicspath{{figures/}}
\hypersetup{colorlinks=true,urlcolor=blue,citecolor=black,linkcolor=black}

\title{Optimization of Workload Mapping and Scheduling in Scaling Heterogeneous HPC Systems: A Systematic Literature Review}

\author{
  \IEEEauthorblockN{Aasish Kumar Sharma\,\orcidlink{0000-0002-7514-2340}}
  \IEEEauthorblockA{%
    Institute of Computer Science\\
    Georg-August-Universit\"at G\"ottingen / GWDG\\
    G\"ottingen, Germany\\
    e-mail: \texttt{aasish-kumar.sharma@gwdg.de}
  }
  \and
  \IEEEauthorblockN{Julian M.\ Kunkel\,\orcidlink{0000-0002-6915-1179}}
  \IEEEauthorblockA{%
    Institute of Computer Science\\
    Georg-August-Universit\"at G\"ottingen / GWDG\\
    G\"ottingen, Germany\\
    e-mail: \texttt{julian.kunkel@gwdg.de}
  }
}

\begin{document}
\maketitle
\pagestyle{empty}
\thispagestyle{empty}

\begin{abstract}
The increasing heterogeneity of High-Performance Computing (HPC) systems,
incorporating Central Processing Units (CPUs), Graphics Processing Units
(GPUs), Field-Programmable Gate Arrays (FPGAs), and accelerators, makes
workload mapping and scheduling a critical optimization challenge.
We present a systematic literature review following the Preferred
Reporting Items for Systematic Reviews and Meta-Analyses (PRISMA)
2020 guidelines, covering 202 studies published from 2017 to 2026,
and classifying optimization techniques for workload mapping and
scheduling in heterogeneous HPC environments.
We identify four solver classes: mathematical programming, heuristics,
metaheuristics, and emerging methods (ML/AI, hybrid, and quantum).
Makespan minimization is the dominant objective and multi-objective
formulations are near-universal; ML/AI approaches have grown from
negligible to dominating recent publications, while quantum methods
are concentrated almost entirely in 2025--2026.
Critical gaps constrain practical relevance: most studies address
scheduling in isolation without joint mapping treatment; evaluations
are predominantly small-scale and simulation-based; real-system
evidence is sparse; and openly reproducible implementations are rare.
These findings signal that the field's pressing challenge is not
algorithmic diversity but deployment-grade evidence: solutions jointly
formulated, validated at production scale, and openly reproducible.
\end{abstract}

\begin{IEEEkeywords}
  systematic literature review; workload mapping; workload scheduling;
  heterogeneous HPC; optimization; scalability; machine learning;
  quantum computing
\end{IEEEkeywords}

\section{Introduction}

Modern High-Performance Computing (HPC) systems combine Central Processing Units (CPUs), Graphics Processing Units (GPUs), Field-Programmable Gate Arrays (FPGAs), and domain-specific
accelerators across tightly coupled clusters, cloud platforms, edge
nodes, and Compute Continuum~\cite{Balouek2019} environments.
This architectural heterogeneity makes workload \emph{mapping}
(assigning tasks to the most suitable processor type) and workload
\emph{scheduling} (determining execution order under dependency and
resource constraints) two of the most consequential optimization
problems in modern HPC.

Both subproblems are NP-hard~\cite{Pinedo2016}, motivating a rich
ecosystem of techniques: exact mathematical programming, constructive
list-scheduling heuristics, evolutionary metaheuristics, deep
reinforcement learning (RL), graph neural networks (GNNs), and, most
recently, quantum optimization~\cite{Sharma2026QUBO}.
Despite this activity, the literature lacks a systematic, up-to-date
review that jointly covers the full solver-class landscape, the
mapping-scheduling distinction, and scalability of evaluated problem
sizes: a dimension critically underexplored relative to its
practical importance in production HPC deployments.

Production HPC workflows in bioinformatics, climate modeling, and
seismic processing routinely involve tens of thousands of tasks across
hundreds of heterogeneous nodes.
Yet the vast majority of published scheduling algorithms are evaluated
on instances of fewer than 1,000 tasks under simulation, abstracting
away system-software overhead and communication latency effects.
This disconnect between research benchmarks and production workloads
is one of the most urgent open gaps in the field.

Our review covers publications from 2017 to 2026.
We restrict the scope to this period to characterize the recent
evolution of heterogeneous scheduling during the time when
accelerator-rich HPC systems and learning-based approaches became
increasingly prominent~\cite{Mao2019,Sharma2025GNN}.
We acknowledge that joint mapping-scheduling approaches exist in the
pre-2017 literature; our scope is deliberately restricted to the
Machine Learning (ML)-era, where the solver-class landscape and scalability demands
differ qualitatively from earlier work.
Prior surveys address narrower subdomains, as shown in
Table~\ref{tab:related_surveys}.
Ahmad et al.~\cite{Ahmad2021} cover container scheduling but not HPC;
Rodriguez and Buyya~\cite{Rodriguez2017} target Infrastructure as a Service (IaaS) cloud exclusively;
Stavrinides and Karatza~\cite{Stavrinides2019} survey cloud workflow
scheduling but do not cover mapping, quantum methods, or provide a
scalability assessment.
To our knowledge, none of the prior surveys jointly addresses the
mapping subproblem, includes quantum methods, and provides a
quantitative scalability assessment of evaluated problem sizes.

We conduct a systematic literature review of 202 studies following
PRISMA 2020 guidelines and contribute:
\begin{itemize}[nosep]
  \item A four-class solver taxonomy (mathematical programming,
        heuristics, metaheuristics, and emerging methods) that
        explicitly names Machine Learning/Artificial Intelligence (ML/AI), hybrid, and quantum as distinct
        sub-classes of emerging methods~\cite{Stavrinides2019,Mao2019,Sharma2026QUBO};
        to our knowledge, the first taxonomy to treat quantum scheduling
        alongside ML/AI in a unified emerging-methods category.
  \item Quantitative analysis across five research questions (RQ1--RQ5)
        with a dedicated scalability assessment.
  \item Identification of the mapping-scheduling gap: only 26.2\% of
        202 studies jointly optimize both subproblems.
  \item A research roadmap for scalable, reproducible, and integrated
        scheduling frameworks for next-generation HPC.
\end{itemize}
The remainder of this paper is organized as follows.
Section~II describes the background and methodology.
Section~III presents the algorithm taxonomy.
Section~IV reports key findings across five research questions.
Section~V discusses implications and a research roadmap.
Section~VI concludes.

\section{Background and Methodology}

\subsection{Problem Definition}

A heterogeneous system has a set of processing elements
$\mathcal{P}=\{p_1,\ldots,p_m\}$ with diverse types
$\tau(p_j)\in\{\mathrm{CPU,\,GPU,\,FPGA,\ldots}\}$.
A workload is modeled as a Directed Acyclic Graph (DAG)
$G=(\mathcal{V},\mathcal{E})$, where nodes $v_i\in\mathcal{V}$
represent tasks and edges $e(i,j)\in\mathcal{E}$ represent data
dependencies with communication volume~$c_{ij}$.

\emph{Mapping} determines the assignment $\mathrm{map}:
\mathcal{V}\!\to\!\mathcal{P}$, subject to feature-compatibility
constraints (a GPU task must land on a GPU node).
\emph{Scheduling} determines start times~$s_i$ satisfying
$s_j \geq s_i + w_{i,\mathrm{map}(v_i)} + c_{ij}\cdot
\mathbf{1}[\mathrm{map}(v_i)\neq\mathrm{map}(v_j)]$,
where $w_{i,j}$ denotes the execution time of task~$v_i$ on
processor~$p_j$.
The canonical objective is makespan minimization:
$C_{\max}=\max_{v_i}(s_i+w_{i,\mathrm{map}(v_i)})$.
The problem $R\,|\,\mathrm{prec},c_{ij}\,|\,C_{\max}$
is strongly NP-hard~\cite{Pinedo2016}, motivating approximate and
learned approaches for realistic scales.

We distinguish mapping (processor-type assignment) from scheduling
(execution ordering) because many scheduling algorithms implicitly
assume a fixed mapping; only studies that explicitly treat both
dimensions are classified here as jointly addressing
mapping and scheduling.
This operational distinction is non-trivial in practice: an algorithm
that selects among $\{$CPU, GPU, FPGA$\}$ types for each task
before determining execution order is mapping-aware, while one that
assumes all processors are of the same type and optimizes only
start-time ordering is not, even if it is called a ``heterogeneous
scheduler'' by its authors.
The 26.2\% figure for joint mapping-scheduling studies reflects
this strict definition and likely overstates prevalence: some
studies use the term ``mapping'' loosely to refer to any task-to-node
assignment without distinguishing processor types.

\subsection{Comparison with Related Surveys}

Table~\ref{tab:related_surveys} positions this review against the three most closely related surveys across four distinguishing dimensions.

\begin{table}[ht]
\centering
\caption{Positioning against existing surveys.
Map.~=~addresses mapping; Q~=~includes quantum; SLR~=~PRISMA
methodology; Scale~=~scalability assessment.}
\label{tab:related_surveys}
\setlength{\tabcolsep}{3pt}
\footnotesize
\begin{tabular}{p{3.5cm}ccccc}
\toprule
\textbf{Survey} & \textbf{Year} & \textbf{Map.} & \textbf{Q} &
  \textbf{SLR} & \textbf{Scale} \\
\midrule
Ahmad et al.~\cite{Ahmad2021}            & 2022 & - & - & \checkmark & - \\
Rodriguez \& Buyya~\cite{Rodriguez2017}  & 2017 & - & - & \checkmark & - \\
Stavrinides \& Karatza~\cite{Stavrinides2019} & 2019 & - & - & - & - \\
\midrule
\textbf{This work}                       & \textbf{2026} & \checkmark
  & \checkmark & \checkmark & \checkmark \\
\bottomrule
\end{tabular}
\end{table}

\subsection{Systematic Review Methodology}

We follow Kitchenham et al.~\cite{Kitchenham2009} and report
according to PRISMA 2020~\cite{Page2021}.
Five thematic queries were executed on Google Scholar (Semantic
Scholar used for metadata enrichment), covering publications from
January 2017 to early 2026; the final search ran on 21 December 2025,
and ten 2026 studies are online-first releases identified at that
time.
After deduplication, 871 unique records were retrieved. The database
search was supplemented by backward and forward snowballing of eligible
studies; this process identified four additional studies that satisfied the
same inclusion criteria.
The full query strings, search logs, and extracted dataset ($N=202$)
are available in the replication package on \href{https://github.com/AasishKumarSharma/slr-hpc-scheduling}{GitHub}.
We address five research questions:
(RQ1)~What optimization techniques are used and how are they
classified?
(RQ2)~What objectives and metrics are targeted?
(RQ3)~What problem characteristics and constraints are addressed?
(RQ4)~What is the state of experimental evaluation, particularly at
scale?
(RQ5)~What are the open challenges and research gaps?

Six inclusion criteria require that each study:
(I1)~addresses mapping or scheduling optimization;
(I2)~targets heterogeneous computing;
(I3)~proposes or evaluates an algorithmic method;
(I4)~is peer-reviewed;
(I5)~falls within 2017--2026;
and (I6)~is published in English.
Exclusion criteria remove homogeneous-only systems, network-only
scheduling, non-peer-reviewed material, and exact duplicates.
Preprints that had not yet been peer-reviewed at the time of
extraction were excluded to maintain the peer-review criterion,
even when a corresponding conference or journal version was
expected; this is acknowledged as a potential source of
underrepresentation for very recent work.

For each included study, 48 data fields were extracted spanning eight
categories: bibliographic metadata, system characteristics (scope:
HPC/cloud/edge/hybrid), workload model, mapping/scheduling treatment,
solver class, optimization objectives, constraint types, and
evaluation practices (simulation vs.\ real-system, maximum task and
resource scale, benchmarks used, and reproducibility indicators
including code availability and dataset links).

Figure~\ref{fig:prisma} shows the PRISMA flow diagram.
After title/abstract screening, 465 of 871 records were excluded for
scope mismatch (homogeneous systems, non-HPC), non-peer-reviewed
status, or duplicates; 406 papers underwent full-text assessment,
of which 208 were excluded (14 related surveys, 5 non-peer-reviewed,
2 duplicates, 187 scope mismatch), yielding 198 studies from the
database search.
Backward and forward snowballing was used as a supplementary search
procedure to identify relevant studies not captured by the database
search. Four additional studies satisfied the eligibility criteria,
giving \emph{202 primary studies} in total (198 from the database
search and 4 identified through snowballing). 

\begin{figure}[t]
\centering
\resizebox{\columnwidth}{!}{%
\begin{tikzpicture}[
  box/.style={rectangle,draw,rounded corners,minimum width=4.0cm,
              minimum height=0.68cm,align=center,font=\scriptsize,
              fill=blue!8},
  excl/.style={rectangle,draw,dashed,minimum width=2.8cm,
               minimum height=0.56cm,align=center,font=\scriptsize\itshape,
               fill=gray!8},
  arr/.style={-{Stealth[length=1.4mm]},thick}
]
\node[box] (id)
  {Identification\\Database records after deduplication: 871};
\node[box,below=0.40cm of id] (scr)
  {Screening\\Title/abstract screened: 871};
\node[excl,right=0.5cm of scr] (ex1)
  {Excluded: 465\\(scope / homogeneous /\\non-peer-reviewed)};
\node[box,below=0.40cm of scr] (elig)
  {Eligibility\\Full-text assessed: 406};
\node[excl,right=0.5cm of elig] (ex2)
  {Excluded: 208\\(14 surveys; 5 non-peer-rev.;\\2 duplicates; 187 scope)};
\node[box,below=0.40cm of elig] (dbincl)
  {Included from database search: \textbf{198}};
\node[box,right=0.55cm of dbincl,fill=blue!8] (snow)
  {Additional eligible studies\\identified through snowballing: \textbf{4}};
\node[box,below=0.48cm of dbincl,fill=green!12,xshift=1.75cm] (incl)
  {Final corpus\\Primary studies: \textbf{202}};
\draw[arr] (id)--(scr);
\draw[arr] (scr)--(elig);
\draw[arr] (scr)--(ex1);
\draw[arr] (elig)--(dbincl);
\draw[arr] (elig)--(ex2);
\draw[arr] (dbincl)--(incl);
\draw[arr] (snow)--(incl);
\end{tikzpicture}%
}
\caption{PRISMA flow for study selection ($N = 202$).}
\label{fig:prisma}
\end{figure}

The primary search database is Google Scholar, supplemented by
Semantic Scholar for metadata enrichment; extending to IEEE Xplore
and ACM Digital Library is planned for the extended journal version.
Data extraction was performed by the first author with supervisory
review by the second author; formal inter-rater reliability metrics
(e.g., Cohen's $\kappa$) are planned for the extended work and are
acknowledged as a limitation of this conference version.
Regarding quantum methods: search queries were progressively refined
to include emerging quantum terminology as the field matured.
The 16 quantum papers from 2025--2026 were independently checked
against the review's eligibility and extraction criteria. Their late
appearance is consistent with recent field growth (quantum
scheduling in HPC gained practical attention only from 2023 onward),
although temporal coverage may also be influenced by progressive
refinement of quantum-specific search terminology.
The preprint inclusion criterion (I4 requiring peer-review) means
that several early-access quantum scheduling papers may have been
excluded if they were not yet formally published; this may slightly
understate the quantum class size.

\section{Algorithm Taxonomy}

\begin{figure}[t]
  \centering
  \includegraphics[width=\columnwidth]{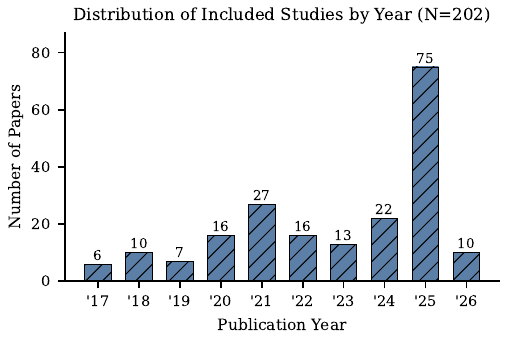}
  \caption{Distribution of 202 included studies by publication year
           (2017--2026). A first wave peaked in 2021 (27 papers);
           a surge in 2025 (75 papers) reflects AI workloads and
           emerging quantum scheduling.}
  \label{fig:pubtrend}
\end{figure}

Figure~\ref{fig:pubtrend} shows the distribution of included studies
by publication year.
Publication activity grew steadily from 2017, reached a first peak
in 2021 (27 papers), dipped in 2022--2023, then surged dramatically
in 2025 (75 papers, the corpus peak), coinciding with increased AI
workloads, accelerator adoption, and the emergence of quantum
scheduling methods.

We derive four top-level solver classes from a bottom-up analysis of
47 distinct algorithms identified across the 202 studies.
Classification follows three criteria: optimality guarantee, solution
construction strategy, and presence of a learning or quantum component.
Table~\ref{tab:taxonomy} summarizes the distribution.

The taxonomy serves two functions: descriptive classification of what
exists in the literature, and prescriptive identification of which
classes are mapping-aware (column Map.).
Mapping awareness is marked as full (\checkmark) when the algorithm
explicitly assigns tasks to processor types as part of its
optimization, or partial ($\sim$) when processor assignment is
implicit or limited to homogeneous-type selection.
The temporal distribution is informative: classical methods (heuristic
and metaheuristic) dominate the pre-2020 literature, while Emerging
Methods (ML/AI, hybrid, and quantum) dominate post-2020 and show
the strongest growth trajectory within the review period.
We use the term \emph{Emerging Methods} to denote solver sub-classes that
arose or gained significant traction within the 2017--2026 review
horizon~\cite{Stavrinides2019,Mao2019,Sharma2025GNN,Sharma2026QUBO},
relative to the classical established methods that predate this
period~\cite{Topcuoglu2002,Talbi2009,Pinedo2016}; the label reflects
historical trajectory within the review window, not current maturity.

\begin{table}[t]
\centering
\caption{Four-class solver taxonomy: classification criteria and
representative algorithms ($N=202$). Map.~=~mapping-aware.}
\label{tab:taxonomy}
\setlength{\tabcolsep}{2pt}
\footnotesize
\begin{tabular}{p{1.9cm}p{1.6cm}p{2.2cm}c}
\toprule
\textbf{Class} & \textbf{Criteria} & \textbf{Examples} &
  \textbf{Map.} \\
\midrule
Math.\ Programming & Optimality guarantee
                   & MILP, CP-SAT, B\&B       & \checkmark \\
Heuristic          & Rule-based; no learning
                   & HEFT, CPOP, EFT          & $\sim$ \\
Metaheuristic      & Stochastic; no learning
                   & GA, DE, PSO, ACO, NSGA-II & $\sim$ \\
Emerging Methods   & Data-driven or quantum:  & DRL, GNN+PPO,
                                               QUBO, QAOA,   & \checkmark \\
\quad ML/AI        & \quad learns from data   & RL+MILP        & \\
\quad Hybrid       & \quad combines $\geq$2   & GA+HEFT        & \\
\quad Quantum      & \quad quantum circuits   & VQE, QAOA      & \\
\bottomrule
\end{tabular}
\end{table}

\emph{Mathematical Programming} methods, including Mixed-Integer Linear Programming (MILP), Constraint Programming with the Boolean Satisfiability solver (CP-SAT), and branch-and-bound, can provide provably optimal solutions. In the studies reviewed, these methods were predominantly evaluated on instances below 100 tasks, reflecting practical tractability limits for the full NP-hard formulation without decomposition~\cite{Pinedo2016}.
Their primary role is as oracle baselines and for solving small
structured subproblems within decomposition frameworks, or for
validating heuristic quality on instances where optimality is
verifiable.

\emph{Heuristics} use problem-specific rules to build complete
schedules deterministically and efficiently.
Heterogeneous Earliest Finish Time (HEFT)~\cite{Topcuoglu2002}, using upward-rank task prioritization and
earliest-finish-time processor selection, is the single
most-referenced algorithm in the corpus, appearing as the primary
method or baseline in 15 studies.
Variants such as Critical Path on a Processor (CPOP), Predict Earliest Finish Time (PEFT), and Modified Predict Earliest Finish Time (MPEFT) extend HEFT with look-ahead and multi-level priority mechanisms.

\emph{Metaheuristics} stochastically explore the solution space
beyond local optima using population- or trajectory-based
search~\cite{Talbi2009,Blum2003}.
Evolutionary algorithms dominate: Genetic Algorithm (GA) and Differential Evolution (DE)
are the most frequent representatives.
Swarm intelligence methods, including Particle Swarm Optimization (PSO), Ant Colony Optimization (ACO), and Artificial Bee Colony (ABC), also appear frequently. Multi-objective variants, such as the Non-dominated Sorting Genetic Algorithm II (NSGA-II) and Multi-Objective Evolutionary Algorithm based on Decomposition (MOEA/D), address joint makespan, energy, and cost optimization.

\emph{Emerging Methods} subsume three interconnected solver sub-classes: Machine Learning/Artificial Intelligence (ML/AI), Hybrid, and Quantum methods.
\emph{ML/AI} methods learn scheduling policies from data or experience:
deep Reinforcement Learning (RL) agents observe Direct Acyclic Graph (DAG) state and learn assignment decisions via
reward signals; Graph Neural Network (GNN)-based schedulers~\cite{Sharma2025GNN,Mao2019}
encode task graph topology into node embeddings for topology-aware
generalization.
\emph{Hybrid} methods combine two or more classes to exploit
complementary strengths: seeding a metaheuristic (GA, PSO) with a
HEFT solution to accelerate convergence, using MILP for mapping and
list-scheduling for ordering, or embedding a learned RL policy as a
local search operator within a metaheuristic.
\emph{Quantum} methods apply Quadratic Unconstrained Binary Optimization (QUBO) formulations and variational quantum algorithms, including the Variational Quantum Eigensolver (VQE) and Quantum Approximate Optimization Algorithm (QAOA), to scheduling subproblems~\cite{Sharma2026QUBO};
16 of 19 quantum papers appeared in 2025--2026, the highest
two-year concentration in the review period.
Because many studies combine multiple approaches, all class counts
use an any-presence (multi-label) scheme: a paper is counted in
every class it employs. Quantitative distributions are reported
under RQ1 (Section~\ref{sec:findings}).

\section{Key Findings (RQ1--RQ5)}
\label{sec:findings}

\subsection{RQ1: Solver Class Distribution and Temporal Trends}

\begin{figure}[t]
  \centering
  \includegraphics[width=\columnwidth]{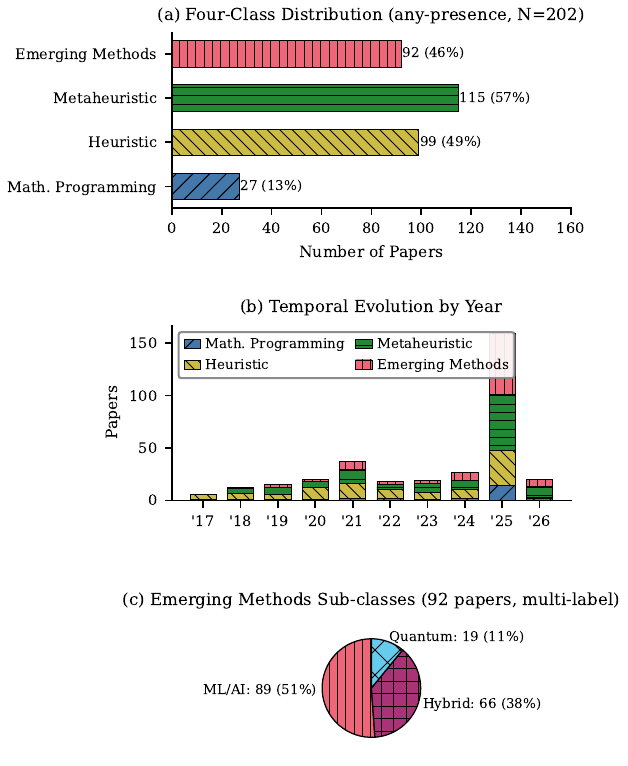}
  \caption{Four-class solver taxonomy across 202 studies.
           (a)~Distribution (any-presence, multi-label). (b)~Temporal
           evolution: Emerging Methods grew from 0\% (2017) to 77.3\% of
           annual papers (2025). (c)~Sub-class composition of the
           92 Emerging Methods papers (multi-label; counts and
           within-sub-class shares shown in chart).}
  \label{fig:solver}
\end{figure}

Figure~\ref{fig:solver} shows the four-class distribution and
temporal evolution.
Mathematical programming appears in 13.4\% of studies (27/202),
heuristics in 49.0\% (99/202), metaheuristics in 56.9\% (115/202),
and Emerging Methods in 45.5\% (92/202).
Within Emerging Methods: ML/AI 44.1\% (89/202),
hybrid approaches 32.7\% (66/202), and quantum 9.4\% (19/202).
The temporal breakdown reveals a major structural shift: Emerging
Methods grew from \emph{0\%} of annual publications in 2017 to
\emph{77.3\%} in 2025, driven primarily by ML/AI adoption.

\begin{figure}[t]
  \centering
  \includegraphics[width=\columnwidth]{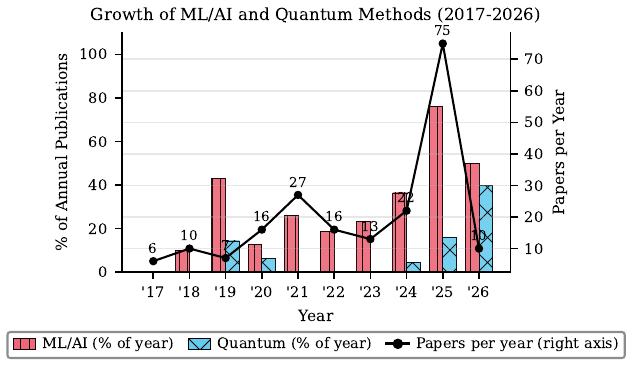}
  \caption{Growth of ML/AI-based scheduling approaches (2017--2026).
           The sharp rise from 2020 reflects maturation of deep RL
           and GNN-based schedulers.}
  \label{fig:mltrend}
\end{figure}

Figure~\ref{fig:mltrend} isolates the ML/AI growth trend.
The inflection around 2020 coincides with the maturation of deep RL
frameworks such as Proximal Policy Optimization (PPO) and Soft Actor-Critic (SAC) and GNN-based combinatorial optimization
libraries that lowered the barrier to applying learned methods to
combinatorial scheduling problems.
Quantum scheduling represents a rapidly emerging sub-class with
19 studies (9.4\%); 16 of these 19 papers appeared in 2025--2026,
the highest concentration in any two-year window of the review period.
The parallel emergence of ML/AI dominance and quantum scheduling
suggests that coming years of HPC scheduling research may be
increasingly shaped by hybrid approaches that combine classical
heuristics with learned or quantum components, rather than by any
single solver class.

The Compute Continuum~\cite{Balouek2019} (the seamless integration
of edge, cloud, and HPC resources) creates additional pressure on
scheduling algorithms to be heterogeneity-aware at multiple scales.
A workflow may begin on edge nodes for data collection, move to cloud
for preprocessing, and execute compute-intensive kernels on HPC
accelerators.
Only a joint mapping-scheduling solver that reasons about processor
types across all tiers can optimize such workflows end-to-end, yet
no study in the corpus addresses this multi-tier scheduling scenario
explicitly.
This gap between the Compute Continuum architecture and the
single-cluster focus of current scheduling research represents an
additional open challenge for the community.

\subsection{RQ2: Optimization Objectives}

Makespan ($C_{\max}$) dominates at \emph{88.1\%} of studies,
confirming its central role as the primary performance metric.
Multi-objective formulations appear in \emph{79.2\%} of studies;
the most common objective pair is makespan + energy
(co-occurring in 71.9\% of multi-objective studies), followed by
makespan + reliability (70.0\%) and makespan + deadline (69.4\%).
Energy (78.2\%), reliability (76.2\%), deadline/Quality of Service (QoS) (75.2\%), and
cost (74.8\%) are the most frequent secondary objectives.
We note that our coding counted a metric as an objective only when it
was explicitly optimized or minimized; metrics that were only evaluated
or discussed after a makespan-only optimization were not classified as
objectives, to avoid overestimating the prevalence of multi-objective
formulations.
Solver classes show distinct objective profiles: heuristic methods
target makespan predominantly alone, while metaheuristic and hybrid
methods more frequently address multi-objective formulations.
This reflects the natural fit of population-based search for
Pareto-front approximation: NSGA-II and MOEA/D natively produce
solution sets that trade off makespan against energy or cost,
whereas single-solution heuristics require explicit multi-objective
extensions (MOHEFT, MOPEFT) to address more than one criterion.
The underrepresentation of energy objectives in Mathematical
Programming methods reflects a structural modeling challenge:
energy-aware formulations can introduce additional power-state
variables and piecewise-linear or nonlinear power relationships,
increasing model size and solution complexity.
This may partly explain the greater use of metaheuristic and ML/AI
methods for energy-aware multi-objective scheduling, where
population-based search naturally produces Pareto-front
approximations across makespan and energy objectives.

\subsection{RQ3: Problem Characteristics and the Mapping-Scheduling Gap}

\emph{Workload models.}
DAG-based workflow models appear in \emph{86.6\%} of studies,
reflecting the prevalence of scientific pipeline workloads
(bioinformatics, astrophysics, seismic processing) in HPC.
Batch job scheduling (independent jobs with resource requests) accounts
for most of the remainder and is more prevalent in pure HPC-cluster
settings. Emerging workload types such as Deep Neural Network (DNN) training pipelines, containerized microservices, and real-time streaming are underrepresented,
representing an important direction for future work.
All 202 included studies target heterogeneous platforms (by inclusion
criterion I2); 97.0\% were explicitly confirmed heterogeneous from
their abstract or methodology text, with the remaining 3.0\%
confirmed through experimental setup tables and evaluation
environments rather than direct textual statements.

\emph{The mapping-scheduling gap.}
A structural bias pervades the literature: \emph{72.8\%} of studies
address execution-order scheduling only, while only
\emph{26.2\%} jointly optimize mapping and scheduling.
In heterogeneous systems, however, the mapping decision (which
processor type handles which task) is often the dominant factor
in solution quality: execution times can differ by an order of
magnitude or more across CPU, GPU, and FPGA resources.
Our operational definition of joint treatment requires explicit
handling of processor-type assignment as a distinct optimization
dimension; many scheduling algorithms implicitly perform a form
of assignment, but are not classified as mapping-aware here.
This gap represents one of the most consequential open challenges
in the literature.
The underrepresentation of joint mapping-scheduling is compounded by
a benchmark artifact: the dominant Pegasus workflow dataset provides
execution time estimates for homogeneous resources, and studies that
use it without modification must artificially simulate heterogeneous
execution times via fixed speed-ratio scaling factors.
This circumvents the fundamental mapping challenge, in which the
processor-type assignment decision determines qualitatively different
execution characteristics, not merely scaled versions of the same
runtime.
Benchmark datasets with native heterogeneous execution profiles
(real GPU/FPGA runtimes per task) are needed to properly expose the
mapping subproblem and prevent this systematic undercount.

Communication cost ($c_{ij}$) is the most frequently modeled
constraint (75.7\%, 153/202), reflecting the prevalence of distributed
execution where inter-node data transfer dominates total makespan.
Feature compatibility constraints (restricting task-to-processor
assignment by hardware type, e.g., GPU tasks must run on GPU nodes)
are explicitly modeled in \emph{35.1\%} of studies (71/202).
Memory constraints appear in 50.0\% of studies (101/202);
security and fault-tolerance constraints are rare.
While 35.1\% is not negligible, it means that the majority of studies
do not formally enforce hardware-type compatibility even while
targeting heterogeneous platforms: most define heterogeneity through
differing execution speeds ($w_{i,j}$ matrices) rather than explicit
type-restriction constraints, a distinction with major practical
implications for real GPU/FPGA HPC deployments.

\subsection{RQ4: Scalability: The Central Challenge}

\begin{figure}[t]
  \centering
  \includegraphics[width=\columnwidth]{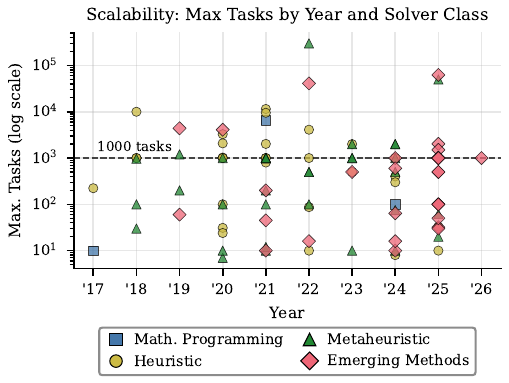}
  \caption{Maximum evaluated task count by four-class taxonomy and year
           (101 of 202 studies report a numeric count; 101 do not).
           Among reporting studies, 78\% evaluate at $\leq$1,000 tasks;
           only 5 exceed 10,000 tasks (maximum: 300,000 in 2022).}
  \label{fig:scalability}
\end{figure}

Figure~\ref{fig:scalability} plots the maximum task count evaluated
against publication year, colored by four-class taxonomy.
The most consequential finding of this review is that the field
faces a severe and persistent scalability ceiling.

\emph{Instance scale.}
Of the 202 studies, 101 (50.0\%) do not report a maximum evaluated
task count at all; the remaining 101 report numeric values.
Among those 101, 79 (78.2\%) evaluate at 1,000 tasks or fewer and
22 (21.8\%) exceed 1,000.
Only 5 studies exceed 10,000 tasks; the largest evaluated 300,000
tasks in a 2022 cloud-resource-management study.
In the reporting subset, Mathematical Programming evaluations are
concentrated below 100 tasks; heuristic and metaheuristic studies
predominantly use 100--1,000 tasks; the five studies above 10,000
tasks all belong to the Emerging Methods class.
Only one study (300,000 tasks, 2022) reaches beyond 100,000 tasks;
none approaches the scale of the largest production HPC batch workloads.
The gap between what is evaluated and what production systems require
is not a minor discrepancy: a 1,000-task study may show a 10\%
improvement over HEFT, but whether this persists at $10^5$-task
scale on a real heterogeneous cluster remains unknown.
The four classes differ in their scalability profiles: Mathematical
Programming methods face tractability limits below 100 tasks in the
reviewed literature; metaheuristics degrade gracefully but require
careful parameter tuning at scale; within Emerging Methods, ML/AI approaches
require large training sets difficult to obtain for realistic HPC workloads,
and quantum methods are constrained by current hardware qubit counts
(typically 20--50 for near-term devices), with the number of schedulable
tasks depending on QUBO encoding density and generally in the tens for
full exact solutions.

\emph{Evaluation method.}
Only \emph{5.0\%} of studies rely exclusively on real hardware;
a further 45.0\% combine simulation with at least one real-system
experiment, but simulation remains the primary evaluation vehicle.
CloudSim, WorkflowSim, and custom event-driven simulators dominate.
Simulation enables large-scale parameter sweeps but introduces
fidelity gaps: simplified communication models, idealized resource
behavior, and no system-software overhead.
The absence of real-system evaluation is particularly concerning for
ML/AI methods, where the sim-to-real transfer problem is well-known:
a policy trained on a simulated environment may fail to generalize
to the stochastic behavior of real heterogeneous hardware,
network contention, and OS-level interference.

\emph{Benchmark diversity.}
Pegasus scientific workflows, including Montage, CyberShake, Epigenomics, SIPHT, and the Laser Interferometer Gravitational-Wave Observatory (LIGO) workflow, are used in the majority of workflow scheduling studies.
Synthetic random DAGs with parameterized task count and
Communication-to-Computation Ratio (CCR) are the second most common
input, but parameter ranges are rarely standardized, limiting
cross-study comparison.
Large-scale HPC job-trace benchmarks, such as the RIKEN Integrated Cluster of Clusters (RICC), San Diego Supercomputer Center (SDSC) Blue Horizon, and Lawrence Livermore National Laboratory (LLNL) Atlas traces, appear in only a small fraction of studies.
To our knowledge, no community benchmark suite for heterogeneous HPC
workflow scheduling currently exists, making it difficult to compare
results across studies.
A meaningful heterogeneous HPC benchmark suite would need four
properties absent from current benchmarks.
First, processor-type-specific execution times measured on real hardware
for at least CPU, GPU, and FPGA node types, not synthetic speed-ratio
proxies applied to a single execution-time matrix.
Second, parameterized problem instances spanning 100 to 100,000+ tasks,
so that scalability profiles can be characterized across the full range
relevant to the field rather than at a single arbitrarily chosen size.
Third, communication overhead models derived from actual interconnect
measurements, including NVLink, PCIe, and InfiniBand transfer latencies,
rather than simple data-volume multipliers.
Fourth, a standardized baseline comparison using HEFT executed on
the same hardware configuration, enabling normalized makespan
improvement ratios that are comparable across studies.
The Pegasus workflow suite, while valuable for cloud simulation, was
designed before GPU/FPGA accelerators became prevalent in HPC;
its homogeneous execution assumptions make it structurally incapable of
exposing the mapping subproblem, which is precisely the gap this review
identifies as most consequential.

\emph{Reproducibility.}
Open-source code or data is explicitly confirmed in only \emph{3.5\%}
of studies (7/202; 57/202 explicitly withhold code; 138/202 make no statement).
Non-standardized inputs, closed-source implementations, and simulation-only
evaluation mean that performance rankings across methods are largely incomparable.
This reproducibility gap is especially consequential for ML/AI methods,
where training hyperparameters, random seeds, and dataset splits
critically affect reported performance.
The gap is starkly counterproductive: ML/AI techniques appear in
44.1\% of all included studies and grew to 76\% of 2025 annual
publications, yet the overall open-source rate (3.5\%) has not kept
pace, preventing independent verification of whether reported gains
over classical heuristics hold across input distributions, system
configurations, and implementation choices.
Minimum reproducibility standards for ML/AI scheduling papers
should include: the trained model weights or a reproducible training
script, the exact DAG instances and execution-time matrices used
for evaluation, and the HEFT baseline run under identical simulator
settings to enable fair comparison.
Without these artifacts, performance claims cannot be independently
verified, and the literature risks accumulating a large body of
incomparable results on favorable synthetic instances.

\subsection{RQ5: Open Challenges}

Six overarching challenges emerge from the quantitative analysis.
First, \emph{integrated mapping-scheduling} remains unsolved: the
26.2\% of studies that address both subproblems typically use
decomposition or sequential approaches rather than joint formulations.
Fully joint optimization is computationally harder but necessary in
truly heterogeneous environments where processor-type choice and
execution order are interdependent.

Second, \emph{production-scale validation} is critically limited:
among the 101 studies that report a task count, only 5 exceed
10,000 tasks, and only one reaches the $10^5$ scale (300,000 tasks, 2022);
none approaches the $10^6$ scale of the largest production HPC batch workloads.
The gap between the 1,000-task evaluation norm and production-scale
workloads involving tens of thousands of tasks across hundreds of
heterogeneous nodes represents one of the most consequential
unsolved empirical problems in the field.

Third, \emph{feature-compatibility-aware scheduling} for GPU/FPGA
targets is formally modeled in only 35.1\% of studies, meaning the
majority do not enforce hardware-type restrictions despite targeting
heterogeneous clusters where not all tasks can run on all node types.

Fourth, \emph{dynamic and online scheduling}, meaning adaptation to
resource failures, job arrivals, and heterogeneous load, is addressed in
72.8\% of studies but rarely validated with realistic trace-driven
workloads.
Most dynamic evaluations use synthetic arrival processes rather than
real HPC job submission logs.

Fifth, \emph{integration with production schedulers} (Slurm,
Kubernetes, PBS) remains limited to a small number of prototypes.
A method that improves makespan by 20\% in simulation but requires
a full scheduler replacement has no practical adoption pathway;
plugin-based integration is required for real-world impact.

Sixth, \emph{domain and workload generalization} across application
types remains unaddressed.
The corpus concentrates on scientific DAG-structured workflows
(bioinformatics, astrophysics, seismic processing), where inter-task
dependencies are stable, execution-time distributions are relatively
narrow, and communication patterns are determined at compile time.
Deep learning training pipelines, containerized microservices, and
real-time streaming workloads exhibit fundamentally different
characteristics: dynamic graph structures, highly variable task
durations, and time-varying resource demands that invalidate
compile-time task-to-processor mappings.
We did not identify a study in the reviewed corpus that evaluates
its scheduler on more than one application-domain class or quantifies
cross-domain generalization.
As HPC systems increasingly co-schedule diverse workload types on
shared heterogeneous infrastructure, cross-domain evaluation
becomes a necessary, and currently missing, validity criterion.

\section{Discussion and Research Roadmap}

The findings reveal a consistent pattern: research effort has
concentrated on small-scale, simulation-only evaluation of isolated
scheduling, while simultaneously undergoing a paradigm shift toward
learned and quantum optimization that could, in principle, scale
better than classical heuristics.
Bridging this gap requires four high-priority directions.

\emph{Joint mapping-scheduling at scale.}
Co-optimizing task-to-processor assignment and execution ordering
remains an important open direction.
The 26.2\% mapping-scheduling gap exists because the two subproblems
are typically solved independently: a mapping step assigns task types
to processor types, followed by a scheduling step that determines
start times.
Joint formulations expose the interaction: the optimal execution order
is inseparable from the processor assignment, especially when
communication costs between heterogeneous nodes dominate.
Candidate approaches include decomposition frameworks, scalable
heuristics, learning-based methods, and hybrid combinations.
Integrating such solvers into workflow engine plugins (e.g., a
Snakemake executor~\cite{Molder2021}) provides a deployable path
to real-system validation.

\emph{Standardized scalability benchmarks.}
The field lacks shared benchmarks for heterogeneous scheduling
analogous to IO500 for storage performance.
Community-curated task graphs with heterogeneous compute-time
matrices, parameterized CCR,
and realistic dependency structures are needed for reproducible
cross-study comparison.
To our knowledge, no community benchmark suite for heterogeneous HPC
workflow scheduling currently exists; the dominant benchmarks (Pegasus
workflows, synthetic DAGs) were designed for cloud simulation
environments, not GPU/FPGA-aware HPC clusters.
Such benchmarks would also enable systematic evaluation of the
mapping-scheduling gap across solver classes.
Community releases under open licenses with persistent identifiers
(e.g., Zenodo DOIs) would simultaneously incentivize reproducibility:
algorithm submissions benchmarked on shared open instances must
release their implementations to be independently replicable.

\emph{Quantum optimization integration.}
QUBO formulations~\cite{Sharma2026QUBO} enable both quantum hardware
and classical quantum-inspired algorithms (simulated annealing,
coherent Ising machines) to attack scheduling subproblems.
As illustrated by the authors' own feasibility study~\cite{Sharma2026QUBO}
(external to the systematic corpus), QUBO encoding of the mapping
subproblem is tractable for up to 20--30 tasks on current near-term
devices; as quantum hardware scales in qubit count and coherence time,
hybrid classical-quantum schedulers may become competitive for larger
structured mapping problems.
The emergence of 16 quantum scheduling papers in 2025--2026
(from zero before 2023) signals that this is an active and fast-moving
frontier where standardized benchmarks are especially needed for
fair comparison.

\emph{Real-system integration and reproducibility.}
Methods should be embedded into production schedulers with feedback
loops for online adaptation.
Workflow engine plugins provide a practical path for end-to-end
validation on real heterogeneous HPC infrastructure.
Releasing open-source implementations, instance datasets, and
environment specifications archived via Zenodo with persistent
identifiers should become standard practice, not the exception.
Multi-fidelity evaluation pipelines, combining fast simulation
screening with real-system spot-validation on a representative
instance subset, allow scalability assessment without prohibitive
hardware cost.

\section{Conclusion and Future Work}

This systematic literature review of 202 studies (2017--2026),
conducted following PRISMA 2020 guidelines, provides a quantitative
account of optimization techniques for workload mapping and
scheduling in heterogeneous HPC systems, with scalability as the
central analytical lens.

The four-class solver taxonomy reveals three structural findings.
First, the solver landscape is dominated by Emerging Methods (ML/AI,
hybrid, and quantum), present in 45.5\% of studies (92/202) and
growing from \emph{0\%} in 2017 to \emph{77.3\%} of annual publications
in 2025. Within this class, ML/AI appears in 44.1\% of all studies,
and quantum scheduling in 9.4\% (19/202), with 16 of 19 papers
appearing in 2025--2026, the highest two-year concentration.
Second, the mapping-scheduling gap persists: only \emph{26.2\%} of
studies jointly address both subproblems.
Third, evaluation practices have not kept pace with algorithmic
growth. Half of all studies omit task-count reporting; among the
101 that do, 78\% evaluate at 1,000 tasks or fewer and only one
(300,000 tasks, 2022) exceeds 100,000 tasks, far below production-HPC
workload scales. Real-hardware exclusive evaluation covers only
\emph{5.0\%} of the corpus, and open-source implementations are
confirmed in just \emph{3.5\%}.

These findings reveal a fundamental imbalance: algorithmic coverage
is broad and rapidly growing, but deployment-grade evidence is
critically absent.
Simulation-only benchmarks, closed-source implementations, and
isolated scheduling formulations remain the norm a decade into the
deep learning era: further algorithmic innovation should be accompanied
by deployment-scale validation, joint mapping-scheduling formulations,
and openly reproducible implementations.

The comparison with prior surveys (Table~\ref{tab:related_surveys}) further clarifies the scope of this review. Stavrinides and Karatza~\cite{Stavrinides2019} examine heterogeneous scheduling, but quantum methods, explicit mapping analysis, and quantitative scalability assessment are not covered together. Rodriguez and Buyya~\cite{Rodriguez2017} focus on IaaS cloud scheduling, while Ahmad et al.~\cite{Ahmad2021} focus on container orchestration. In contrast, this review analyzes mapping and scheduling across heterogeneous HPC environments using a common taxonomy spanning exact, heuristic, metaheuristic, ML/AI, hybrid, and quantum methods, together with quantitative analysis of evaluation scale.

The identified gaps motivate four directions for further research: stronger integration of mapping and scheduling, evaluation at larger and more representative workload scales, wider release of implementations and experimental artifacts, and increased validation on real heterogeneous systems and production scheduling environments.

\section*{Acknowledgment}

The authors acknowledge the North-German Supercomputing Alliance
(GWDG / NHR@G\"{o}ttingen) for providing computational resources
used in the scheduling experiments referenced in this review.
This work is part of the DECICE project, funded under the European
Commission Horizon Europe programme.

\printbibliography

\end{document}